# Hall effect evidence for an interplay between electronic correlations and Na order induced electronic bands topology in Na$_x$CoO$_2$

I.F. Gilmutdinov[1*], I.R. Mukhamedshin[1,2] and H. Alloul[2]

[1] Institute of Physics, Kazan Federal University, 420008 Kazan, Russia,

[2] Université Paris-Saclay, CNRS, Laboratoire de Physique des Solides, 91405, Orsay, France.

## Abstract

The incidence of topology on the band structure and physical properties of layered compounds has been extensively studied in semimetals. How those evolve in presence of electronic correlations has been less investigated so far. In the sodium cobaltates Na$_x$CoO$_2$ considered here, unexpected magnetic properties associated with correlations on the Co sites have been disclosed about 15 years ago. Altogether it has also been found that various orderings of the Na atoms occur in between the CoO$_2$ layers in the stable phases of these cobaltates. The distinct Na orders of these phases have been shown to induce specific Co charge disproportionation with large size unit cells in the CoO$_2$ planes, linked with the diverse magnetic behaviors. This provides an original playground in the studies of interplays between topology and correlations in these layered materials. We present here transport measurements on a series of single crystals and demonstrate that we do synthetize pure phases with quite reproducible transport properties. We show that above room $T$ those display a similar behavior whatever the Na content. On the contrary we provide evidence for a great diversity in Hall effect low temperature dependences which underlines the specificities of the Fermi Surface reconstructions induced by the Na order. We study in some detail the difference between two metallic phases, one ($x = 0.77$) antiferromagnetic below $T_N = 22$ K and the second ($x = 2/3$) paramagnetic down to $T = 0$. Both show a sign change in the Hall effect with decreasing $T$. We demonstrate that this can be attributed to quite distinct physical effects. For the $x = 0.77$ phase the negative Hall effect has a non-linear field dependence and occurs in similar conditions to the anomalous Hall effect found in various magnetic metals. In the $x = 2/3$ phase in which the Co sites are disproportionated in a kagome substructure, the negative sign of the Hall effect and its two steps $T$ variation have to be assigned to specificities of its Fermi surface. In both cases the anomalies detected in the Hall effect are certainly associated with the topology of their Co electronic bands.

[*] ildar.gilmutdinov@gmail.com; present address: Laboratoire National des Champs Magnétiques Intenses, 31400 Toulouse (France)

# I  INTRODUCTION

A great evolution in the physical studies of layered metals has occurred in the last 30 years. The discovery of the high-$T_C$ superconductivity (SC) in cuprates has been at the origin of many research activities on transition metal oxide materials. The importance of electronic correlations (EC) on the Cu sites has been underlined by the existence of their magnetic properties (1). Studies on other transition metal compounds have been mostly motivated by the search for unconventional SC properties. For instance a special attention has been devoted to the cobaltates $Na_xCoO_2$ after the discovery of SC with $T_C \sim 5$ K in a hydrated phase for $x \approx 0.3$ (2). In these compounds the Co sites form a triangular two dimension (2D) lattice and magnetism was unexpectedly found to occur for $x > 0.7$ (3) (4). The interest in these compounds faded away as SC has not been found so far in non-hydrated compounds.

Later on the discovery of the properties of the graphene layer (5) with Dirac cones in its band structure (6) has aroused a great interest in materials for which the electronic bands involve singular points protected by their topology (7). Specific materials often with unit cells without inversion symmetry have been shown to exhibit such singular Dirac or Weyl (8) points (the latter occur above or below the Fermi level). Those have been initially evidenced in materials with weak electronic correlations as is the case for the single-layer graphene. But nowadays it has becomes clear that such topological properties might occur as well in materials with strong EC and can yield original magnetic properties which might become useful in spintronics applications. There is therefore a growing interest in correlated electron compounds in which topological effects might occur.

Coming back to the cobaltates which are clearly correlated electron systems, their triangular lattice structure differentiates them from the nearly square lattice cuprate structure. Furthermore nuclear magnetic and quadrupolar resonance (NMR/NQR) experiments on aligned powder samples allowed to disclose a series of Na ordered phases (9) also seen by transmission electron microscopy (10). The local spin susceptibilities measured by the Co NMR shift allowed to reveal specific disproportionations of the Co charges with singular large size unit cells in the $CoO_2$ layers. For instance, a kagome Co sublattice has been shown to exist for $x = 2/3$ (11) (12). We thus point out that EC and topological states are concomitant in these materials. This underlines the interest to consider in some detail the influence of the charge order on the Co sites on the physical properties in real and k spaces.

So far the sodium cobaltates have been found metallic for nearly any Na content. But the initial experiments did not reveal a clear picture of the actual evolution of the band structure with the Na concentration. Calculations within a local density approximation (LDA) ignoring the Na order do not reproduce the charge differentiation. They imply a large hexagonal 2D Fermi surface (FS) with eventually some small pockets for low Na doping (13) (14). Those results were apparently partly supported by angle-resolved photoemission spectroscopy data (ARPES) although small pockets were never observed (15). On the contrary the quantum oscillations (QO) detected in initial transport measurements imply the existence of small carrier pockets in the ground state (16).

Some care has to be exercised in the analyses of those initial experiments. ARPES is a technique which probes surface states. After cleaving the samples between $CoO_2$ planes in the ab direction the average Na content left on the top layer should be $x/2$ in order to satisfy charge neutrality. This cannot maintain the same ordered state of the Co planes as the bulk. On the contrary transport experiments are quite sensitive to the Fermi surface reconstructions induced by the Co charge order in the bulk single crystals. But in that respect the main difficulty has been so far the production of homogeneous and reproducible single crystals. We have done recently specific efforts to control as much as possible the quality of single crystal samples of ordered Na phases (17). This enables us now to probe the bulk electronic properties by resistivity and Hall effect measurements. The high quality of our samples is reflected by the large differences found between the various ordered phases and by the reproducibility of our data. We specifically study here the $T$ and applied magnetic field B dependences of the Hall effect which is currently a good probe of unusual incidences of the topology on the band structure of metals and semi-metals.

The paper is organized along the following line. We shall first display in Sec. II the large differences in transport we did evidence on a series of well controlled ordered Na phases. This permits us to reveal in many cases sign changes of the Hall effect which evidences that the Co disproportionation induced by the Na order induces a multiband character. We select then two phases which apparently display at low $T$ similar large negative values of the Hall constant. In Sec.III we demonstrate that in the $x = 0.77$ phase (18) this behavior is connected with the establishment of an ordered metallic antiferromagnetic (AF) state at low $T$. On the contrary we show in Sec. IV that for the $x = 2/3$ phase (11) the sign change of the Hall effect occurs in a purely paramagnetic state and must be associated with the specific band structure of this phase. A rough quantitative analysis of the Hall effect performed in Sec. V leads us to suggest that a single band picture might apply for the AF phase while multiband behavior is required to explain the Hall effect in the $x = 2/3$ phase. We also discuss the possible analogies of the anomalous behavior of the Hall effect (AHE) compared to that seen in various magnetic metals. Similarly, we consider the possible relation between the Hall anomalies seen in the $x = 2/3$ compound and its reconstructed kagome band structure.

## II. COMPARISON OF TRANSPORT PROPERTIES OF DIFFERENT PHASES

We have performed transport measurements with the standard techniques detailed in Appendix A. The longitudinal resistance $R_{xx}$ and the transverse resistance $R_{xy}$ measured in an applied field $B$ give respective measures of the resistivity $\rho$ and of the Hall constant $R_H$ after taking into account the geometrical factors of the sample. Data have been taken on single crystals of five phases which had been singled out using powder samples (19). The distinct Na orders revealed by NMR in some of these phases are illustrated in Fig. 1(a). We found that some characteristics of the resistivity and Hall effect data were perceptible in the initial papers. (20)

*Resistivity data*. As can be seen in the upper Fig. 1(b), all pure phases were found to display metallic resistivity except that for $x = 1/2$. This phase had been identified early on and is known to exhibit a magnetic transition at 86 K (21) (22) not detected by transport techniques and a metal insulator transition at $T \approx 51\text{-}53$ K (23). The latter is clearly seen in

Fig. 1(c), in which a secondary increase of $\rho$ is seen to occur below 20 K, so that the resistivity increases by a factor ~100 from 50 K to low $T$.

For the other phases we measured in our single crystals residual resistivities $\rho_0$ much lower than those found in former publications (with resistivity ratios as high as a few hundred given in Appendix A). The $\rho_0$ values and the differences in the quoted Na content indicate that the samples studied in the early papers were quite often mixtures of Na ordered phases (20).

For all phases, the high-$T$ behaviors of $\rho(T)$ are found quite similar as can be seen in the upper Fig. 1(b). The actual differences in $\rho(T)$ for these phases are mostly seen at low $T$ on the resistivity derivatives $d\rho/dT$ as will be displayed in next sections.

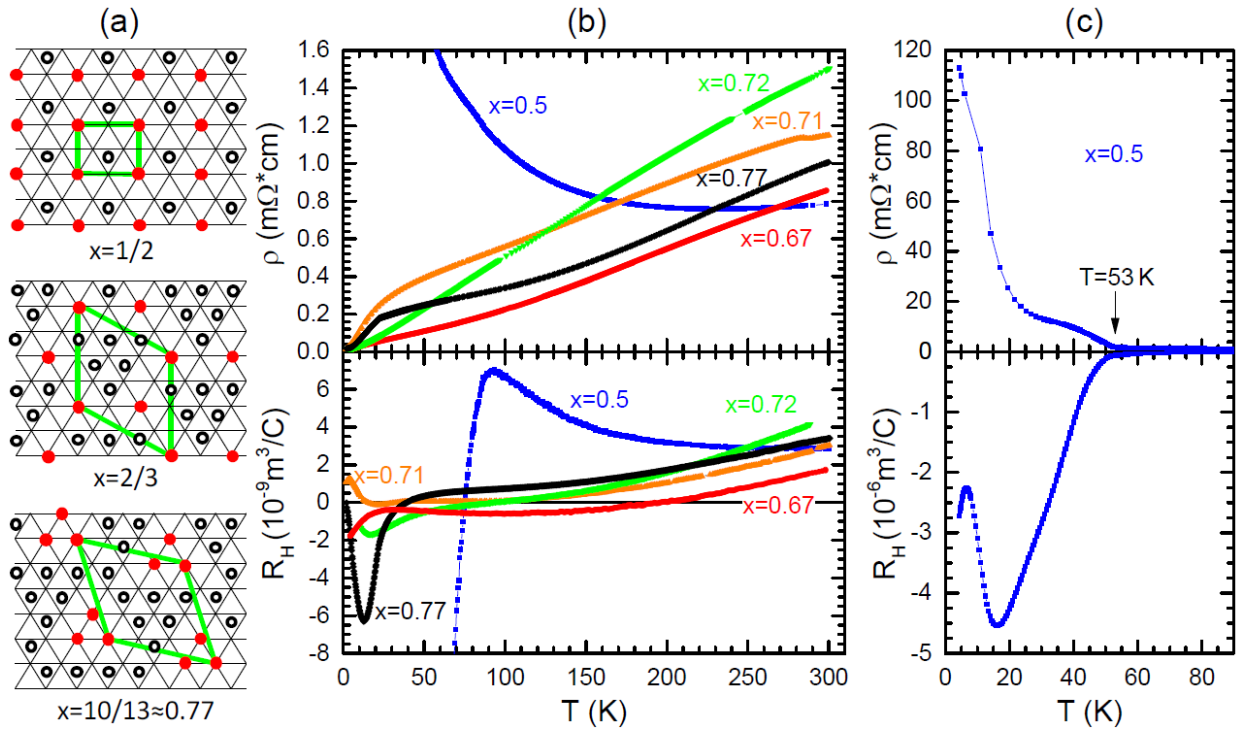

FIG. 1. (Color online) (a) Three typical Na ordered phases. Here the Co atoms located on the triangular lattice sites are not represented. The Na atoms are located either (red) on top of a Co site or (black) on top of a triangle of Co atoms (9). (b) Zero field resistivity and Hall constant data measured in a 9 tesla applied field (respectively in the upper and lower frames) for five typical phases. They differ markedly at low temperatures (see text). (c) The low $T$ data for the $x = 1/2$ phase are displayed on enlarged scales. This underlines the metal to "insulator" transition at 53 K.

*Hall effect*: In these five phases the $T$ dependence of the Hall effect has been found to display major differences. The Hall constant $R_H$ is seen in the lower Fig. 1(b) to become negative in most cases at low temperature. Such a behavior is totally unexpected within the frame used to analyze the initial experiments and suggests that the Fermi surface is $T$ dependent. The $x = 1$ parent phase has been established as being a band insulator involving only $Co^{3+}$ states in which the $t_{2g}$ sublevels are filled (24) (25). From LDA band structure calculations one expected a progressive depletion of these levels with decreasing x. This continuous increase of hole carrier density due to Na vacancy doping is apparently seen for the surface states probed by ARPES measurements (15). One would then expect a $T$-independent positive Hall effect for any Na content. This is never seen to happen in our set of data although a positive Hall effect is always seen above room $T$. The large decrease of the Hall effect and its sign changes which occur with decreasing $T$ are clearly demonstrating that the simple LDA picture does not apply for the planar ordered electronic states.

The negative Hall effect regimes reveal the occurrence of carrier bands with a dominant electronic character and imply that the Na order has induced a Fermi surface reconstruction. This is somewhat similar to the observations done in the underdoped cuprate pseudogap regime (26). This analogy with the present results is then in line with the early detection in the cobaltates of QO due to small Fermi pockets (16) prior to the discovery of QO in the cuprates (27). The NMR (9) and x-ray (12) experiments have given evidence that the Na atomic order is already established at room $T$. So the low $T$ Hall data are certainly due to fine effects of the reconstructed FS topology.

Above room $T$ the positive Hall effect measured in all samples apparently implies that in that $T$ range the transport properties are governed by hole carriers. On another hand $R_H$ is seen to increase regularly with increasing $T$ around room $T$ in all samples. This behavior was clearly noticed even up to much higher $T$ in the initial experiments on a given sample (28). Such a linear increase of the Hall effect has indeed been suggested to be a specific behavior for a uniformly doped triangular metallic lattice, whatever the doping and the EC strength (29) (30). The present results do support quite generally this theoretical suggestion.

The main idea is that this effect is associated with thermal shifts of the chemical potential. A similar suggestion had been proposed from transport (31) experiments in Fe pnictides, and supported by ARPES experiments (32) although the origins of the shift of chemical potential are quite distinct in those compounds. Here it is mainly associated with the triangular structure of the lattice. Let us notice that this $T$ variation is specifically seen in the regime for which Na diffusion and ionic conductivity becomes significant. Does that induce a situation similar to that which prevails for the ARPES data taken on disordered surface states? That could mean that the transport properties become above room temperature insensitive to the Na order induced reconstruction of the FS. We notice however that $R_H$ does not evolve regularly with decreasing x at a given $T$ which could mean that local Na order still has an influence on the slope of the high $T$ variation of $R_H$.

In any case the sign changes of the Hall effect in the lower Fig. 1(b) are always found to occur below room $T$ where the Na order and charge differentiation are already fully established. This happens together with the onset of anomalous magnetic responses that signal the importance of electronic correlations.

The $x = 0.77$ phase which develops an ordered magnetism below $T_N = 22$ K has a positive Hall effect down to low $T \sim 50$ K. But it becomes large and negative when $T$ approaches $T_N$ and *apparently recovers a nearly null value at $T = 0$* (see lower Fig. 1(b)). On the contrary for $x = 2/3$ the Hall effect becomes negative for $T < 200$ K. The magnitude of the Hall effect increases then down to $T = 0$ in two steps in this peculiar phase in which a dis-proportionated state with a kagome like structure occurs. The clearcut difference between these phases justifies our first-step choice here to investigate the actual origin of these distinctly singular behaviors.

## III  AF PHASE WITH $x = 0.77$

Let us consider the most concentrated Na phase which exhibits a definite AF order. In most previous magnetic studies it has been found that an A type AF order is established for most phases with x > 0.75 (33). This corresponds to Co ferromagnetic layers with moments in the $c$ direction which alternate in magnetic orientation from layer to layer.

The AF transition is perfectly apparent in the transport properties as a sharp drop of $\rho(T)$ occurs at $T_N = 22$ K (Fig. 2 (a)) .This is very clear on $d\rho/dT$ (Fig. 2(b)) and quite reproducible as can be seen in Appendix B. The resistivity above $T_N$ is therefore associated with a large magnetic scattering contribution in the paramagnetic (PM) state which is quenched when the magnetization freezes so that a very good metallicity is restored in the low $T$ state. We also see there that the magnetoresistance (MR) is positive and large (≈100%) at $T<<T_N$ and becomes quite negligible in the PM state.

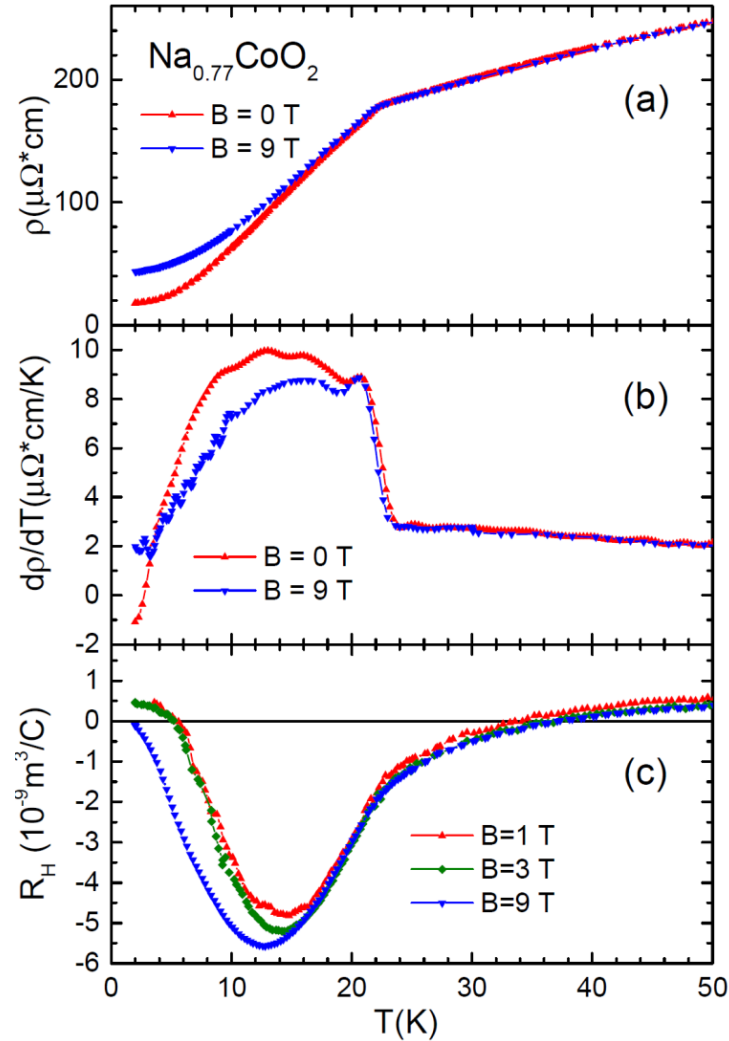

FIG. 2. (Color online) (a) In any applied field the resistivity of the $x = 0.77$ phase drops below $T_N = 22$ K. A quite large positive magnetoresistance appears below $T_N$. (b) The abrupt increase of the derivative of the resistivity at $T_N$ illustrates the sharpness of the AF transition. (c) The Hall constant measured for $B = 9\,T$ is small and positive at high $T$ but progressively becomes large and negative already above $T_N$. For lower applied field a pronounced variation of $R_H$ with B is seen below about 15 K

As for the Hall effect, as can be seen in the lower Fig. 1(b), it has a behaviour quite distinct from that observed in most other phases. First, it remains positive down to 50 K, which is a much lower temperature than in the other phases. Below 50 K it becomes negative and goes through a large maximum in absolute value around 15 K which is found to be quite reproducible (Appendix B). Finally RH decreases towards zero at low T, as emphasized in the low T plot of Fig. 2(c), which is somewhat unique when compared to the other phases. However, we noticed, as shown there, that the Hall constant exhibits also an unexpected field dependence below about 15 K.

The fact that the Hall voltage becomes non linear with the applied field B could be better seen in the data taken at fixed $T$ versus $B$ presented in Fig. 3. We clearly see there that the Hall effect even changes sign with applied field at low temperatures. The simplest analysis of these data can be done by extracting the initial linear in field contribution. The latter is found to be positive at the lowest temperature $T \ll T_N$. But for increasing $T$ this positive contribution is overwhelmed by a negative term which has a more complex $T$ and $B$ dependence. Finally as can be seen in Appendix C on the data taken in a PPMS the non linearity of the field dependence of this negative contribution decreases markedly with increasing $T$ above 15 K so that $R_H$ becomes field independent above $T_N$ = 22 K.

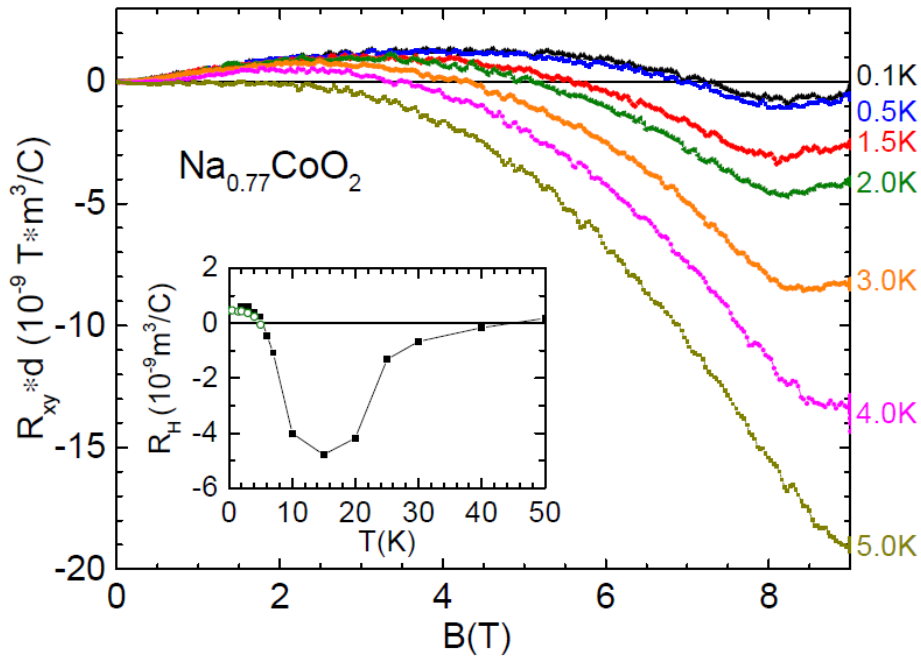

FIG. 3. (Color online) The field dependence of the Hall effect illustrated here below $T = 5$ K gives evidence for two competing contributions. In the insert the linear contribution to the Hall effect extracted from the initial slope is plotted versus temperature. This has been done independently for two samples (empty green circles from the present data and filled squares from the data shown in Appendix C, taken above 2 K in a PPMS on a distinct sample).

We also evidence at low $T$ a kink in the Hall effect at about 8 tesla (both on Fig. 3 and Fig. 8 in Appendix C) which we could associate with a step-like increase seen at the same field on the magnetization (Fig. 9 in Appendix D). This reveals a spin flop or a metamagnetic transition. Therefore all these results clearly clearly evidence that the magnetism has an influence on the scattering of the carriers and induces a magnetoresistance. Such effects appear even above $T_N$ while the static magnetic order is not established and could be still connected with the appearance of a dynamic short range 2D ferromagnetic order. Such anomalous contributions to the Hall effect (AHE) are commonly detected nowadays in diverse metallic magnetic materials and are associated with quite distinct possible origins, as will be discussed later in sec V. In the $x$ = 0.77 phase this contribution apparently disappears in the $T$ = 0 limit at low fields so that the positive linear term might be associated *with the normal Hall contribution of the free carriers*.

## IV   $x$ = 2/3 KAGOME LIKE PHASE

Let us consider now the $x$ = 2/3 phase in which the charge disproportionation has been studied thoroughly by NMR/NQR experiments which did revealed a kagome sub-lattice of Co sites (11) (12). In this phase a large regular decrease of $\rho(T)$ is observed with the smallest residual resistivity found in our five phases, as can be seen in Fig. 1(b) and in Appendix A. Here again the Hall effect plotted in the lower Fig. 1(b) is positive at room $T$ as for the other phases, but It becomes already negative below 200 K. This sign change has been found to be quite reproducible on many samples as can be seen in Fig. 4.

At even lower $T$ of about 40 K a second reproducible Hall effect anomaly is detected. Contrary to the case of the $x$ = 0.77 phase $R_H$ remains negative and even increases in absolute value down to the lowest temperatures. Furthermore, the Hall voltage is found to be linear in field down to the lowest $T$ as displayed in Fig. 5.

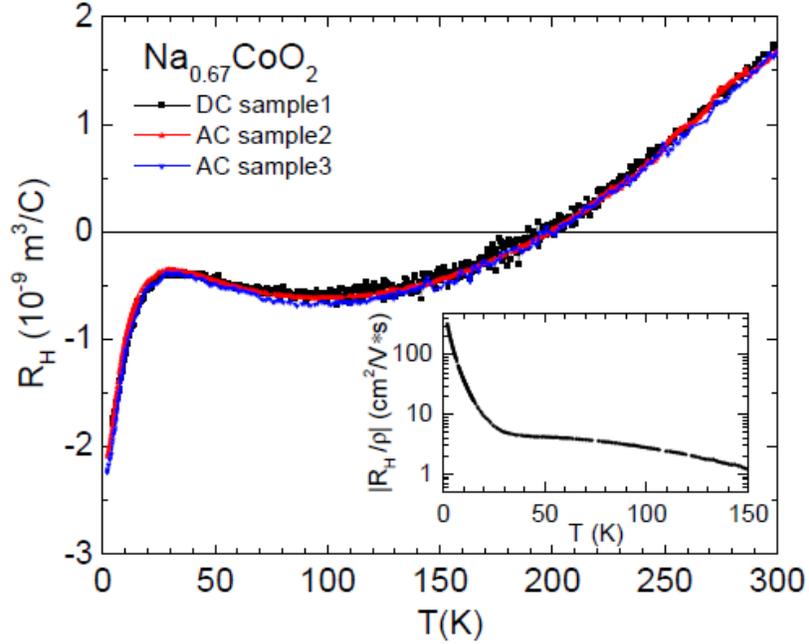

FIG. 4. (Color online) Hall constant for the $x = 2/3$ phase taken in an applied field $B = 9\,T$ in distinct experimental setups, using AC or DC current excitations (see Appendix A). The change of sign of the Hall effect at 200 K and the minimum in its absolute value at 30 K are found to be quite reproducible. In the inset the ratio $|R_H/\rho|$ which would corespond the electron mobility in the single-band model discussed in Sec. V, is reported up to 150 K .

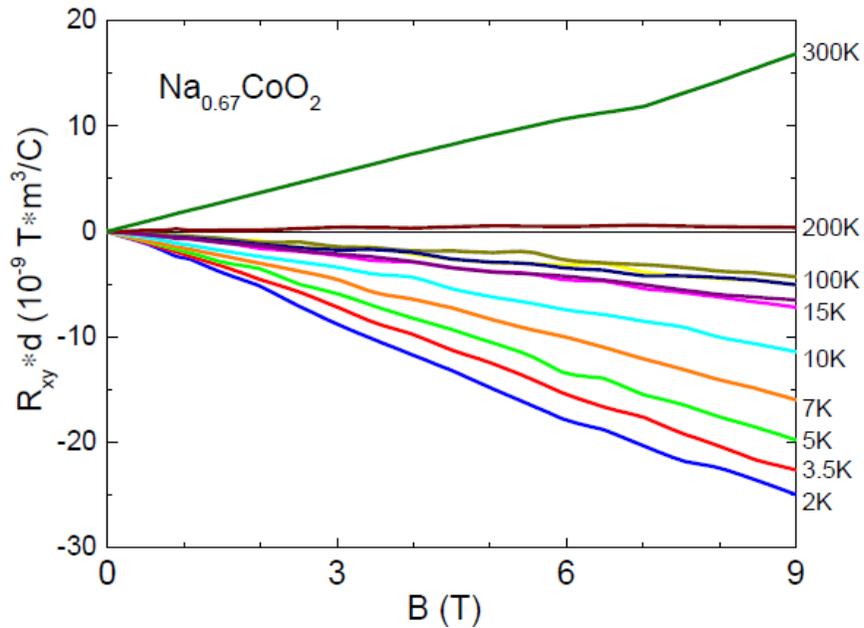

FIG. 5. (Color online) Field dependence of $R_{xy}$ in a $x = 2/3$ sample taken in a PPMS. The linear fits of these data have been used to derive the $R_H$ values reported in the inset of Fig. 3.

If one compares the behavior of the Hall effect with that of the resistivity one can see on Fig. 6 that the onset of the low $T$ decrease of the Hall effect coincides with the onset of a decrease of $d\rho/dT$ which indicates an inflection point at about 30 K which is also found to be quite reproducible (17). We might notice as well that the change of behavior of the Hall effect is concomitant with the large increase of the spin susceptibility detected by NMR without any onset of magnetic order down to $T = 0$ (34).

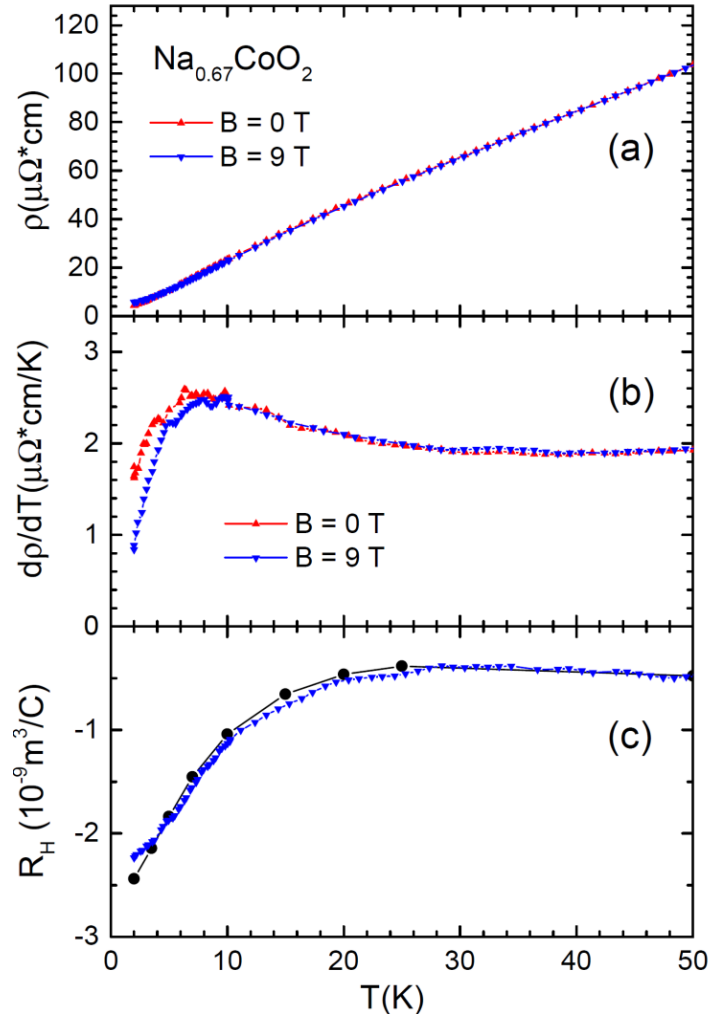

FIG. 6. (Color online) The low-$T$ data measured in zero field for $\rho$ in (a) and $d\rho/dT$ in (b) are compared to those measured for $B = 9$ T. It can be seen that the magnetoresistance remains quite small. The Hall coefficient $R_H$ data measured at 9 T on sample 3 of Fig. 4 are plotted in (c). They display a minimum in absolute value which coincides with the decrease of $d\rho/dT$. These $R_H$ data agree perfectly with those (full circles) obtained on a distinct sample, as deduced from the linear field dependence of the Hall voltage data shown in Fig. 5.

At odds with the *x* = 0.77 data, the resistivity for *B*||*c* does not differ markedly from that measured in zero field. So the magnetoresistance remains small in all the temperature range considered. The large negative contribution to the Hall effect also behaves qualitatively quite differently than for *x* = 0.77. All these facts suggest that the strange behavior of the *T* variation of the Hall effect is linked with narrow features in the metallic band structure and not with the appearance of a magnetic order.

## V  DISCUSSION

We have seen that the negative contribution to the Hall effect seen in the *x* = 0.77 phase appears to be connected with an AHE contribution associated with the magnetic response in the metallic state. Otherwise the normal carrier band contribution to the Hall effect is positive for the *x* = 0.77 phase and negative for the *x* = 2/3 phase. That essentially means that the carriers which dominate the transport properties are holes in the former case and electrons in the latter. Let us consider first the possible origin of these normal contributions to the Hall effect.

Normal charge contribution to the Hall effect

An analysis of the transport properties can be carried out when the Fermi surface is known either by ARPES or at least from LDA calculations as could be done in the case of iron pnictides (31). In such cases for which the FS presents multiple bands and 2D cylindrical carrier pockets with both electronic and hole character, the dominant charge carriers were found somewhat dependent of the doping (35). In the present case as pointed out in the introduction, the reconstructed FS is not known so far and we can at the moment only consider rough models. The simplest one is to assume two types of carriers with respective hole and electron content $n_h$ and $n_e$ and mobility $\mu_h$ and $\mu_e$. In such a case (31) (35)

$$\sigma = \rho^{-1} = n_h e \mu_h + n_e e \mu_e, \quad [1]$$

and

$$R_H \sigma^2 = [n_h e \mu_h^2 - n_e e \mu_e^2] \quad [2]$$

In the $x = 0.77$ phase the leading contribution to $R_H$ at $T = 0$ in low applied field has a magnitude $R_H \approx 0.5 \; 10^{-9} \; m^3/C$ which is rather close to that found at about 50 K in the PM state (insert of Fig. 3). One might then speculate that the normal state positive Hall effect would be $T$ independent below about 50 K in the absence of ordered magnetism. One may even assume that a single band model might apply with $n_e = 0$, which would be compatible with the very small magnetoresistance found experimentally. In such an oversimplified model the mobility would be given by $\mu_h = R_H / \rho$, that is $\mu_h \approx 28 \; cm^2/(V^*sec)$ at $T = 2$ K. This model would result in in a nearly $T$ independent hole content $n_h = (R_H \; e)^{-1} = 0.47 / Co$. Let us notice that this value is twice as large as the 0.23 / Co value expected for a homogeneous doping of the $CoO_2$ planes but would better agree with the disproportionated state value described by the NMR data in this phase (36).

For the $x = 2/3$ phase $R_H$ is always negative below 200 K but displays a singular low $T$ variation. Both hole and electron carrier bands are required. But even if we do apply the charge neutrality condition $n_h - n_e = 1 - x$ we cannot deduce three parameters ($n_e$, $\mu_h$ and $\mu_e$) with the two observables given by equations [1] and [2]. We may however again oversimplify the problem by assuming that $\mu_h << \mu_e$ which allows a reduction of the transport problem to a single electron band case. From the $T = 2$ K value $R_H \approx - 2.4 \; 10^{-9} \; m^3/C$ we would deduce $n_e = |R_H e|^{-1} \approx 0.1 / Co$ which would correspond to a hole content $n_h \approx 0.43 / Co$. But in that case the anomalous decrease in magnitude of the Hall constant $R_H$ with increasing $T$ would correspond to a four-fold increase of the electron carrier content $n_e$ up to 100 K. Also this simple approach would imply the low $T$ anomalous increase of the electron mobility $\mu_e = |R_H/ \rho|$ shown in the insert of Fig. 4. The electron mobility $\mu_e \approx 330 \; cm^2/(V^*sec)$ at $T = 2$ K would be much larger than the hole mobility mentioned above in the 0.77 phase.

In any case it seems clear that high mobility electrons are required to explain those data, but a serious explanation of these results can hardly be obtained without invoking a multi-band behavior with sharp features at the Fermi level. This should not be surprising in view of the large paramagnetism which prevails at low $T$ in this phase, attributed to the Co kagome sublattice demonstrated by NMR data (11). In most calculations done so far for kagome lattices flat bands are obtained. Our results, therefore, call for specific band structure calculations, taking into account the EC and the Na order.

Anomalous contribution to the Hall effect for $x = 0.77$

Let us compare now the AHE detected in the $x = 0.77$ AF phase to the observations done in various magnetic metals. In such cases an AHE has quite often been seen and is somewhat expected for ferromagnetic (37) but not for AF metals. It has however been seen in some specific AF compounds as discussed hereafter. In some cases this AHE has been found dependent on the magnitude of the longitudinal current and named NLHE (for nonlinear HE) (38). This does not apply to the present case as we did find that here the AHE is independent of the current as can be seen in the insert of Fig. 7 in Appendix B. Also an AHE has been observed in AF metals in which the moments do not order with collinear orientations such as cubic $Mn_3Ir$ (39) (40). Another type of AHE is detected even in zero field when a remanent magnetization remains after field cooling. It has been assigned to skyrmions either ordered or pinned by defects (41) and is often named Topological Hall Effect (THE). We do see as well that a remanent magnetization is observed in our samples (see Appendix D). The fact that the negative Hall effect is nonlinear in field in our AF phase, and changes at the spin flop or metamagnetic transition indicates that such possible origins for the AHE might be considered here, and that domain walls might as well play a role. More refined experimental studies of the relation between $R_H$ and the magnetic properties have to be undertaken to determine whether the AHE is connected to local magnetization components which are not aligned along the $c$-axis or might be related to residual disorder.

## VI SUMMARY

We have undertaken studies on various ordered Na phases and shown that the transport data is quite reproducible on our high quality single crystal samples. We disclose major pieces of evidence that the low-temperature transport properties are totally governed by the Co disproportonation associated with the Na ordering. Negative contributions to the Hall effect have been detected and indicate the occurrence of Fermi surface reconstructions similar to the case of cuprates in which a charge order was discovered, and subsequently was found to disrupt the SC order in some cases. The present results therefore reinforce the conclusion that in chemically doped layered materials the planar metallicity is quite generally strongly affected by the chemical order of the dopants as has been suggested initially for the cuprates (42).

We have considered in some detail two phases with low-temperature negative Hall effect contributions. Our results evidence that these observations are related with important problems presently of great interest in the condensed matter community.

In the $x = 0.77$ AF metallic samples we find evidence that hole carriers apparently dominate the 2D charge transport, which seems to be a specific property of this phase. But in the vicinity of the AF state a large negative contribution to the Hall effect is detected and found non-linear in the applied field. We suggest that it has some analogy with the AHE observed in various AF metals. This might be connected with the fact that the Na induced unit cell in the $CoO_2$ planes is non centrosymmetric. Our results raises questions about the possible existence of non collinear contributions to the magnetic order e.g. skyrmions ordered or pinned by defects. Such fundamental questions are often highlighted in the community in particular in view of possible spintronics applications in AF metals with high $T_N$ values. The present results have lead us to plan more detailed studies of the magnetotransport properties in this phase in high field exceeding the metamagnetic transition.

On the contrary, the negative Hall effect is found to be linear in field for the $x = 2/3$ phase in a paramagnetic metallic regime where a disproportionated kagome structure of the $CoO_2$ planes has been established. The detected singular low $T$ variation of the Hall effect can be taken as an evidence for a narrow energy feature in the band structure, such as a flat band which is predicted by many calculations on kagome band structures (43). While kagome lattices are heavily studied in half filled insulating compounds in which a spin liquid state prevails (44) this $x = 2/3$ phase might be considered as a singular and original realization of a hole doped metallic Kagome structure. As QO have been detected initially in some cobaltate samples we are presently engaged in such investigations in our cleaner samples. The attempt to get complementary information on the Fermi surface topology of this pure phase should stimulate specific band structure calculations taking into account the electronic correlations, which are necessary to establish the disproportionated state (45). The size of the unit cell might still be too large for the present state of the art ab initio band structure calculations in correlated electron systems. Model calculations could however help to determine whether Weyl states and Berry phase effects could arise in such experimental conditions. In any case our results raise some original questions about the interplay between topology and correlations in these layered materials.

## ACKNOWLEDGEMENTS

I.G. thanks F. Rullier-Albenque for her initial help in acquiring the experimental expertise required for transport experiments (she unfortunately passed away soon after the start of this project). We acknowledge frequent exchanges with L. Balicas and A. Subeidi, and thank V. Brouet, B. Gaulin and M. Leroux for their careful reading of the manuscript. The crystal growth, XRD, magnetic and transport measurements were carried out at the Federal Center of Shared Facilities of Kazan Federal University. The work of I.R.M. was partially supported by the Russian Science Foundation (Project No. 19-12-00244). Travel between Kazan and Orsay has been financially supported by an "investissement d'avenir" allowance from the Labex PALM (ANR-10-LABX-0039-PALM).

## APPENDIX A : SAMPLES AND MEASUREMENT TECHNIQUES

Single crystals of sodium cobaltates $Na_xCoO_2$ with $x \sim 0.8$ were grown using an optical floating zone technique. Subsequent electrochemical treatment was used in order to decrease sodium content and to obtain crystals with homogeneous sodium distribution. Details of the sample preparation techniques have been described elsewhere (17). X-ray diffraction data allowed us to control the existence of a single value of the *c*-axis lattice parameter that indicates the absence of phase mixing.

The longitudinal and transverse resistances $R_{xx}$ and $R_{xy}$ of the single crystals were measured by the van der Pauw method (46) and in some cases by the conventional six contacts method. Samples with an average size of 2x1 mm were prepared for the conventional resistivity measurements, and square shaped crystals of average size 2x2 mm were prepared for measurements with the van der Pauw method. The sample thickness was less than 100 μm. The gold wires (50 μm diameter) used for the contacts were attached to the samples with silver paint. All resistance measurements were performed with the current in the *ab*-plane of the crystal and magnetic field along the c-axis.

The alternating current transport option (ACT) installed in the physical properties measurements system (PPMS-9, Quantum Design) was used for electrical resistance measurements in the 2-300 K temperature range in applied fields up to 9 tesla. Measurements of the AC electrical resistance at temperatures 0.1-3 K were performed in a dilution refrigerator BF-LD400 (BlueFors Cryogenics) using the same ACT electronics. The AC resistivity measurements frequency was 69 Hz and the current magnitude was usually less than 0.5 mA at temperatures above 2 K and 0.1 mA for lower temperatures.

Hall voltage measurements on samples with six-contact were performed in positive and negative magnetic field (along the c-axis direction), in order to subtract the contribution of the spurious longitudinal voltage due to contacts misalignment. The Hall coefficient was defined as:

$$R_H = \frac{V_{xy}}{I_{xx}} \frac{d}{B} = R_{xy} \frac{d}{B},$$

where $V_{xy}$ is the transverse voltage or Hall voltage, $d$ the sample thickness, $B$ the magnetic field, and $R_{xy}$ is the transverse resistance or Hall resistance.

In Table 1 the resistivity values of the samples from Fig. 1 measured at 300 K and 2 K are given. The residual resistivity of metallic phases is much lower than the values shown in experiments published before [20], and found to be quite reproducible in our experiments. These altogether confirms the high quality of the single crystals used.

Table 1: Measured resistivity at 300 and 2 K for the samples in Fig. 1. The residual resistivity ratio $RRR = \rho_1/\rho_2$ is given for the metallic samples.

| x | $\rho_1$ (T = 300 K) (mΩ*cm) | $\rho_2$ (T = 2 K) (mΩ*cm) | RRR |
|---|---|---|---|
| 0.5 | 0.79 | 113.5 | |
| 0.67 | 0.86 | 0.004 | 215 |
| 0.71 | 1.14 | 0.016 | 71 |
| 0.72 | 1.50 | 0.015 | 100 |
| 0.77 | 1.00 | 0.018 | 56 |

## APPENDIX B: REPRODUCIBILITY OF THE $x = 0.77$ TRANSPORT DATA

Resistivity and Hall coefficient measured for a series of $x = 0.77$ samples are displayed in Fig. 7. The resistivity data have been found to be quite reproducible. The Hall effect measurements have been taken using the van der Pauw technique, after cooling the sample in zero field, then ramping up the field to 9 tesla. The Hall constant $R_H$ was then measured for increasing temperatures in this 9 tesla applied field. The temperature dependences found are quite analogous for these distinct samples, with a large maximal negative values at about 14 K. The small differences in $R_H$ magnitude cannot be assigned to sample geometrical factors. It might be due to slight differences in the remnant magnetization as the actual local magnetization has often an influence on the AHE in AF metals. For a given sample (Cr13-19A Cr1) the data plotted in the insert of Fig. 7 have been taken for longitudinal currents differing by one order of magnitude. This confirms that the Hall effect is linear in the longitudinal current.

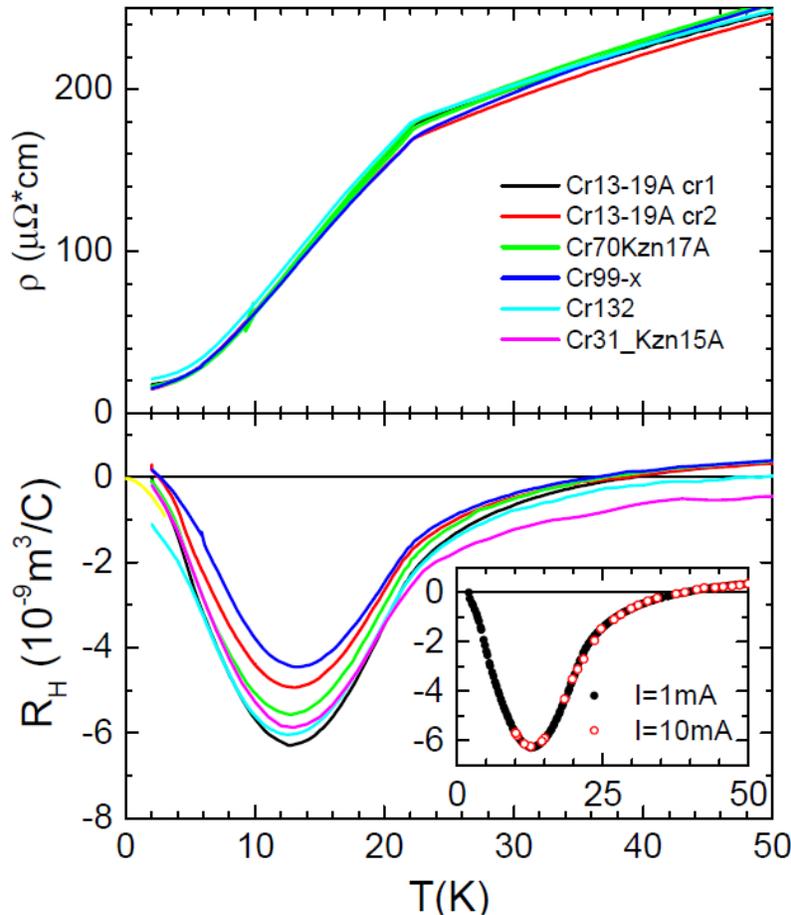

FIG. 7. (Color online) Resistivity (upper frame) and Hall constant (lower frame) taken at low $T$ on a series of 0.77 samples. Insert: $R_H$ data taken for two longitudinal current values on one sample.

## APPENDIX C: FIELD VARIATION OF THE HALL EFFECT FOR $x = 0.77$

For a $x = 0.77$ sample the Hall effect data were taken with the standard six contact method after cooling the sample to the lowest $T$ in zero field. Then $R_{xy}$ vs $B$ curves were measured from the lowest to the highest temperatures – see Fig. 8. The data in this figure are qualitatively analogous to that of Fig. 3 obtained at lower $T$ on a distinct sample with the van der Pauw technique. A linear behavior is recovered at about 15 K. The initial slope analysis of these results has been used to deduce the filled square data points plotted in the inset of Fig. 3.

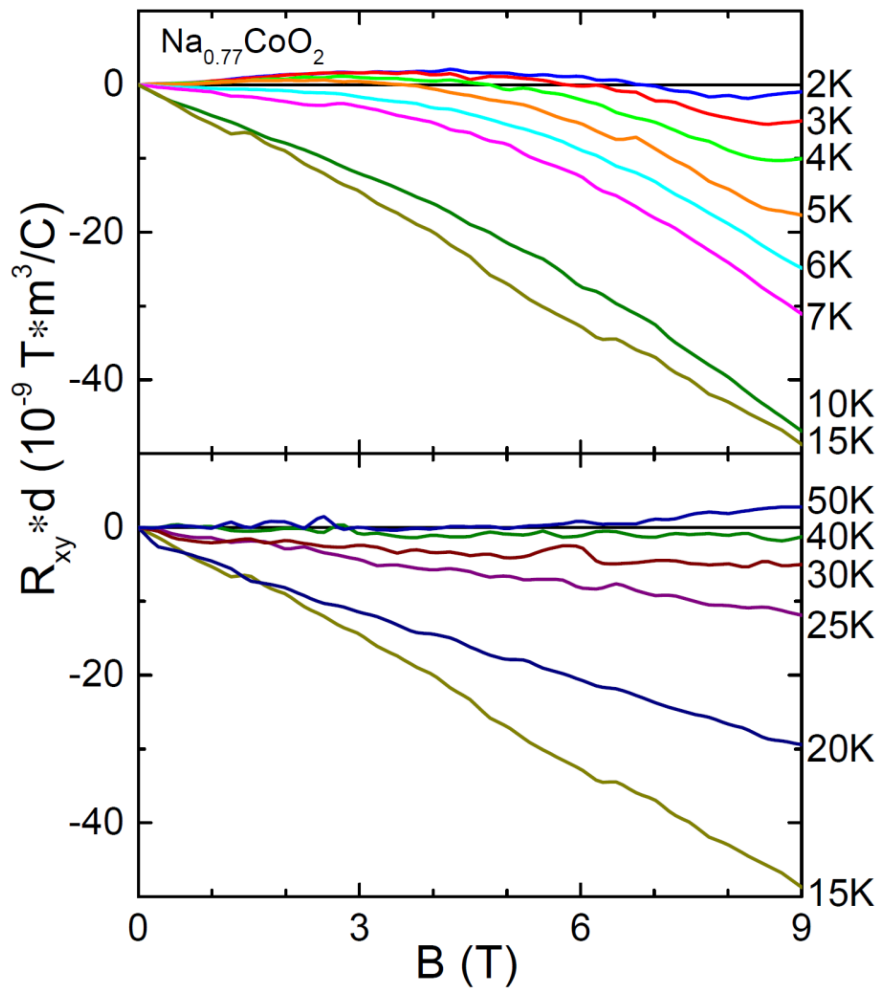

FIG. 8. (Color online) The variation of the Hall constant with applied magnetic field $B$ is plotted for increasing temperature, well below $T_N$ in the upper frame and around and above $T_N$ in the lower frame.

# APPENDIX D: MAGNETIC PROPERTIES OF THE $x = 0.77$ PHASE

The magnetic moment of $x = 0.77$ phase samples was measured using the vibrating sample magnetometer (VSM) of the physical properties measurement system (PPMS, Quantum Design). The magnetic field was applied in the c direction of the crystal. Two types of $T$ dependence of the magnetic moment were measured: the zero-field-cooling (ZFC) curve and the field-cooling curve (FC). At first, the magnetic field was set to zero value in oscillating mode to reduce the remnant magnetic field of the superconducting magnet, then the sample was cooled in zero magnetic field from $T = 300$ K to 2 K, and the magnetic field was ramped to H = 100 Oe in the non-overshoot mode. Then the magnetic moment was measured while warming up (ZFC-curve). The sample was then cooled down back to $T = 2$ K without changing the applied magnetic field, and the magnetic moment was measured while warming up (FC-curve). One can see in this experiment that a remnant magnetic moment is induced in the FC experiment below $T_N = 22$ K, as shown in Fig. 9(a).

After a ZFC the variation of the magnetization was measured in increasing magnetic field. For $T = 25$ K, that is above $T_N$, the $M(B)$ curve of Fig. 9(b) is found linear. At $T = 5$ K in the AF state one can see a step-like increase for $B = 8$ tesla which can be associated with a spin flop of the local moments or a meta-magnetic transition in the case of this metallic magnetic compound. A kink of the Hall constant is also detected for this field value below about 5 K as can be seen in Figs. 3 and 8.

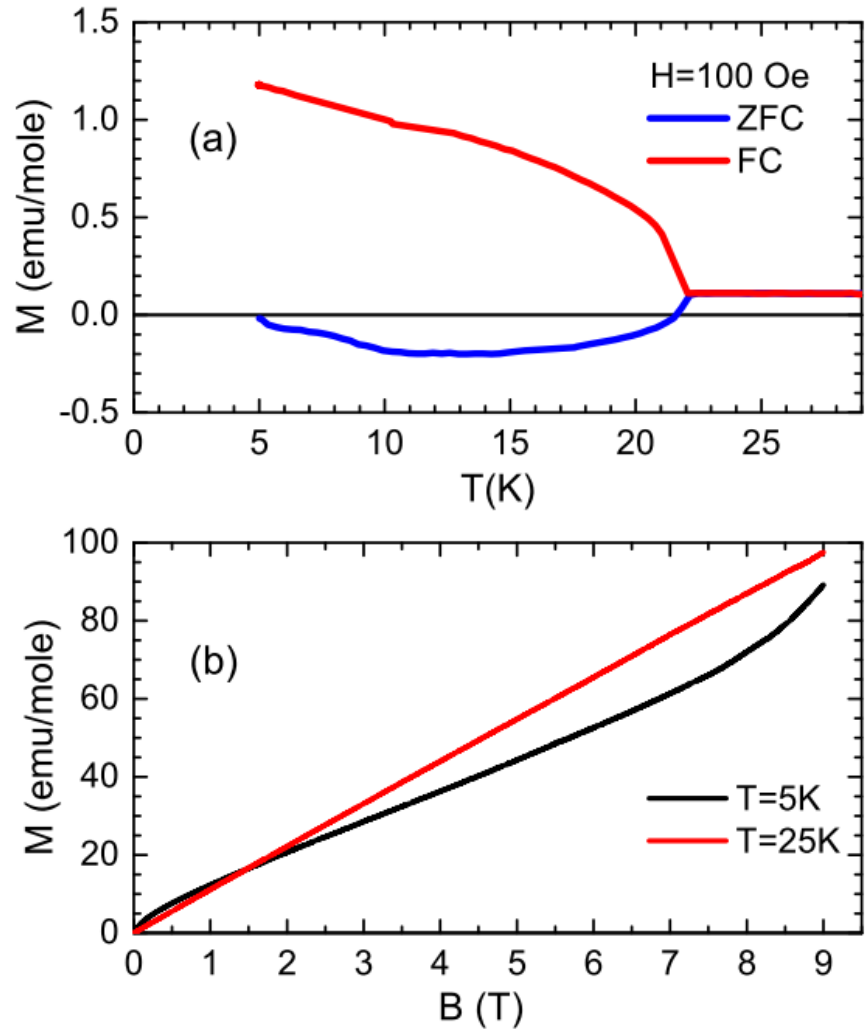

FIG. 9. (Color online) (a) Magnetization measurement on a $x = 0.77$ phase sample after ZFC or FC. (b) Below $T_N$ the field variation of the magnetization after ZFC is shown to display an increase at about 8 $T$ which can be assigned to a spin flop or metamagnetic transition.